\documentstyle[epsfig]{aipproc}

\begin{document}

\def\ang{\AA$\;$}

\title{Quasi-Chemical Theory \\ and \\ Implicit Solvent Models for
Simulations}

\author{Lawrence R. Pratt and Susan B. Rempe}

\address{Theoretical Division, Los Alamos National Laboratory, Los
Alamos, New Mexico 87545 USA }

\maketitle

\begin{abstract}
A statistical thermodynamic development is given of a new implicit
solvent model that avoids the traditional system size limitations of
computer simulation of macromolecular solutions with periodic boundary
conditions.  This implicit solvent model is based upon the
quasi-chemical approach, distinct from the common integral equation
trunk of the theory of liquid solutions.  The idea is geometrically to
define molecular-scale regions attached to the solute macromolecule of
interest.  It is then shown that the quasi-chemical approach
corresponds to calculation of a partition function for an ensemble
analogous to, but not the same as, the grand canonical ensemble for
the solvent in that proximal volume.  The distinctions include: (a)
the defined proximal volume --- the volume of the system that is
treated explicitly --- resides on the solute; (b) the solute
conformational fluctuations are prescribed by statistical
thermodynamics and the proximal volume can fluctuate if the solute
conformation fluctuates; and (c) the interactions of the system with
more distant, extra-system solution species are treated by approximate
physical theories such as dielectric continuum theories.  The theory
makes a definite connection to statistical thermodynamic properties of
the solution and fully dictates volume fluctuations, which can be
awkward in more ambitious approaches.  It is argued that with the
close solvent neighbors treated explicitly the thermodynamic results
become less sensitive to the inevitable approximations in the implicit
solvent theories of the more distant interactions.  The physical
content of this theory is the hypothesis that a small set of solvent
molecules are decisive for these solvation problems.  A detailed
derivation of the quasi-chemical theory escorts the development of
this proposal.  The numerical application of the quasi-chemical
treatment to Li$^+$ ion hydration in liquid water is used to motivate
and exemplify the quasi-chemical theory.  Those results underscore the
fact that the quasi-chemical approach refines the path for utilization
of ion-water cluster results for the statistical thermodynamics of
solutions.

\end{abstract}

\pagebreak

\section*{Introduction}
Direct simulation of macromolecules in aqueous solutions typically
requires consideration of a mass of solution large compared to the
mass of the macromolecular solute.  Frequently, the bulk of the solution
is of secondary interest.  The extravagant allocation of computational
resources for direct treatment of macromolecular solutions limits the
scientific problems that may be tackled.  Thus, {\em implicit solvent}
models that eliminate the direct presence of the solvent in favor of
an approximate description of the solvation effects have received
universal and extended interest
\cite{Berkowitz:82,Brooks:83,Brunger:84,Belch:85,Brooks:85,Brooks:86,%
Pratt:86,Brunger:87,Brooks:89,Brooks:88,Deloof:91,Levchuk:91,%
Shiratori:91,Straub:91,Braatz:92,Tapia:92,Widmalm:92,%
Heller:93,Fritsch:93,Juffer:93,Nakagawa:93,Rashin:93,Beglov:94,%
Arnold:94,Axelsen:94,Davis:94,Lazaridis:94,Pastor:94,Straub:94,%
Sharp:94,Beglov:95,Eriksson:95,Essex:95,Norberg:95,Alaman:96,%
Bencsura:96,David:96,Larwood:96,Luty:96,Wang:96,Albaret:97,%
Haggett:97,Han:97,Meirovitch:97,Parrill:97,Merz:98,Zeng:98,%
Lounnas:99,Prabhu:99,Roux:99a,Roux:99b,Roux:99c,Levy:99}.

These issues are specifically relevant to this workshop on
electrostatic interactions in solution for two reasons.  First,
electrostatic interactions exacerbate the difficulties of solute size
mentioned above.  If all interactions were short-ranged, most
practitioners would be satisfied with the traditional approach,
adopting periodic boundary conditions and empirically examining the
system size dependence of their results by performing calculations on
successively larger systems.  Second, the additional computational
requirements to treat genuinely chemical phenomena in solution by {\it
in situ\/} electronic structure calculations again limits the problems
that can be addressed.  In fact, the principal physical concepts
involved in electronic structure calculations for chemical problems in
liquid water universally involve electrostatic interactions of long
range.  The important example of metal ion chemistry in proteins
combines these points.

Although a variety of implicit solvent models have been tried by now
in numerous selected applications, they are still limited in
fundamental aspects.  A helpful recent discussion of important
limitations was given by Juffer and Berendsen\cite{Juffer:93}.  They
emphasize that implicit solvent models should be {\em significantly}
simpler than explicit models in view of the approximations and {\em ad
hoc} features that are accepted. Solution chemistry problems that
require direct treatment of electronic degrees of freedom are one kind
of problem that must be made significantly simpler to permit a broader
computational attack.  A striking example is the current `ab initio'
molecular dynamics calculations of aqueous solutions of simple ions.
They do treat electronic structure issues within simulations but have
been typically limited to total system sizes of 16 -- 32 water
molecules in periodic boundary conditions\cite{marx:99}, small systems
by current standards with classical simulation models. These small
sizes do limit the conclusions that might be drawn; the structure and
dynamics of the second hydration shell of a Li$^+$ ion in liquid
water, the example discussed below, undoubtedly requires calculation
on systems larger than 32 water molecules. Nevertheless, much can be
learned from the study of such small systems, particularly when
chemical effects of the interactions of a solute with proximal solvent
molecules are the issues of greatest importance\cite{lubin:99}.

The idea for the developments presented here is aggressively to adopt
the Juffer-Berendsen suggestion that implicit solvent models must be
{\em significantly} simpler than explicit models and apply that
philosophy to the statistical thermodynamic treatment of aqueous
solutions.  To that end, we acknowledge that some water molecules play
a specific, almost chemical, role in these hydration phenomena.  We
then work out the theory that permits inclusion of a small number of
such molecules explicitly.  The required theory is a descendent of the
quasi-chemical approximations of Guggenheim\cite{Guggenheim:35},
Bethe\cite{Bethe:35}, and Kikuchi\cite{brush}. It is a significant
simplification of direct simulation calculations. Roughly described,
that theory organizes and justifies treatment of a handful of water
molecules essentially as ligands of the macromolecule of interest,
letting the more distant solution environment be treated by simple
physical approximations such as the popular dielectric continuum
models.

The plan of this presentation is first to introduce an example, the
hydration of the Li$^+$ ion, that permits a convenient discussion of
the quasi-chemical theory.  That example illustrates the
quasi-chemical pattern for the theory, exemplifies the basic molecular
information required to construct quasi-chemical predictions, and
offers a simplified derivation based upon a thermodynamic model. 
Following that we give an extended theoretical development of the
quasi-chemical organization of calculations of solvation free energies
and then use those theoretical results to suggest an explicit-implicit
solvent model for statistical thermodynamic calculations of solutions.

\begin{figure}[b] % fig 0
\centerline{\epsfig{file=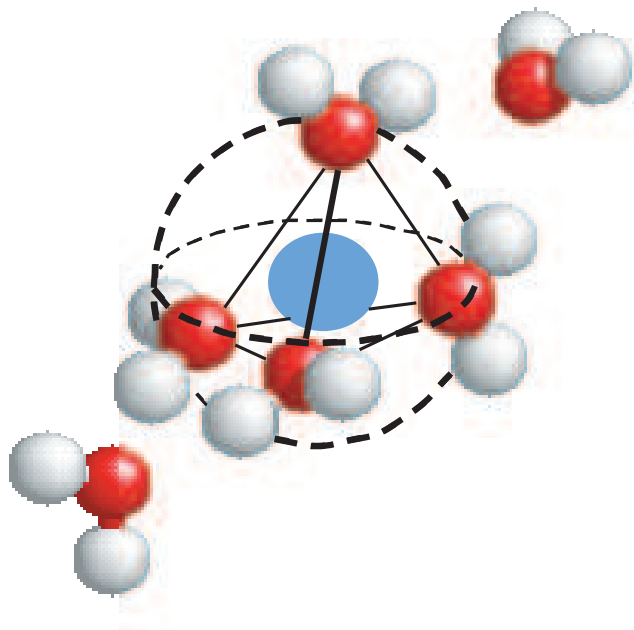,height=3.5in,width=3.5in}}
\vspace{10pt} \caption{The minimum energy structure of a Li$^+$
ion with six water molecules.  The inscribed ball identifies the inner
shell occupied by the four nearest water molecules.  The two
further water molecules are in an outer shell.  This illustrates
the definition of the bonding or inner shell region.}
\label{fig0} \end{figure}
 
\section*{Example: the Li$^+$ Ion in Water}
The hydration of atomic ions provides a conceptually simple context in
which to consider the quasi-chemical approaches developed below.  The
previous study of the hydration of ferric ion, Fe$^{3+}$(aq), provides
one such example\cite{Martin:98}.  Here we motivate the formal
developments of the theory by considering the hydration of the Li$^+$
ion in dilute aqueous solution and we will discuss the principal
theoretical structures in this context first.  We anticipate some of
the discussion below by noting that Li$^+$(aq) proves to be a
difficult case in some important respects.  Thus, we expect to return
to this example in later work.  Indeed, the goal of the theoretical
developments initiated here is of a sufficiently constructive nature
that all pieces of the puzzle needn't become available at the same
time!

\subsection*{Quasi-chemical Structure  for the Hydration Free
Energy} The quasi-chemical theory suggests expressing the chemical
potential of the lithium ion species, $\mu_{Li^+}$, in %the format
terms of ideal and non-ideal, or interaction, contributions:

\begin{eqnarray} &\beta & \mu _{Li^+} = \ln \left[ {{{\rho
_{Li^+}} \over {q_{Li^+}/V}}} \right]-\ln p_0  -  \ln \left[
\sum\limits_{n \ge 0} {\tilde K_{n}}\,\rho _{H_2O}{}^n 
\right], \label{eqostate} \end{eqnarray}

\noindent with $\beta^{-1}=RT$.  
In the last contribution, only a limited number of terms are included
in the sum. That limited number is the maximum number of water
molecules that may occur as inner hydration shell ligands. For the
Li$^+$ ion example, that limited number need not be greater than six
(6). Fig.~\ref{fig0} shows the minimum energy structure of the
Li$^+$ ion with six water molecules\cite{rempe:99}.

The coefficients ${\tilde K_{n}}$ of Eq.~(\ref{eqostate}) will be
defined more fully later. A virtue of this approach is that the
natural initial approximation is ${\tilde K_{n}}\approx K_{n}^{(0)}$,
the equilibrium ratio for the reactions that form various inner shell
clusters

\begin{eqnarray} Li^+\ +\ nH_2O \rightleftharpoons Li^+(H_2O)_n
\label{reaction} \end{eqnarray}

\noindent without consideration of medium effects, as for an ideal
gas. Those coefficients can be assembled within the harmonic
approximation for the cluster vibrations utilizing standard electronic
structure computational packages\cite{gbook}.  That idea is reinforced
by the results in Tables~\ref{T_1}, \ref{T_2}, and \ref{T_3}, which we
use to develop this example\cite{rempe:99}.

\begin{table}

\caption{Electronic energy (E), zero-point energy (ZPE), the
interaction part of the chemical potential ($\mu^*$) for
Li(H$_2$O)$_n{}^+$ complexes, and the Mulliken charge for the Li$^+$ ion
in each complex utilizing the B3LYP and dielectric continuum solvation
approximations.  Geometries and ChelpG partial charges were obtained
with basis 6-31++G** for O, H, and 6-31G* for Li$^+$.  The Mulliken
partial charges were obtained with the smaller basis 6-31G on all
atoms in order to simplify their consideration.  Dielectric radii for
all atoms were taken from 
Ref.~\protect\cite{Stefanovich} except R$_{Li^+}$=2.0~\AA. }

\label{T_1}

\begin{tabular}{lddddd}& E (au) & ZPE (kcal/mol) & $\mu^*$
(kcal/mol)&$\delta$ (e)\\ \hline 
H$_2$O & -76.4332&  13.4 & -8.3 &-\\ 
Li$^+$ &-7.2845  &- &-82.&1.0 \\ 
Li(H$_2$O)$^+$ &-83.7750  &15.5 &-73.&0.9\\
Li(H$_2$O)$_2^+$ &-160.2596  &31.0 &-68.&0.7\\ 
Li(H$_2$O)$_3^+$ &-236.7308 &46.5 &-65.&0.6\\ 
Li(H$_2$O)$_4^+$ &-313.1912 & 61.4 &-65.&0.5\\ 
Li(H$_2$O)$_5^+$ &-389.6390 &76.7&-63.&0.4\\ 
Li(H$_2$O)$_6^+$ &-466.0878 & 93.0 &-61.&0.3\\ 
\hline \end{tabular} \end{table}

Within the quasi-chemical approximation of collective
phenomena\cite{brush}, the factors of solvent density that appear
explicitly in Eq.~(\ref{eqostate}) can be recognized as mean-field
estimates of the influence of the solvent on the chemical potential of
the Li$^+$(aq) ion.  That form of the solvent mean field, however,
derives from specific compositional effects. For the realistic
circumstances that long-ranged interactions of the classic
electrostatic type are important, the approximation ${\tilde
K_{n}}\approx K_{n}^{(0)}$ should be revised with implicit solvent
theories intended to describe the influence of the more distant
solution environment on the chemical potential of the Li$^+$(aq) ion.
This hints at the strategy for the quasi-chemical approaches.
Neighbors occupying a carefully defined proximal volume are treated
explicitly.  But more distant neighbors are treated implicitly with
the view that they can be satisfactorily treated with simple theories
and, in any case, those more distant effects are a smaller part of the
whole.

\begin{table}

\caption{Ideal gas thermochemistries (kcal/mole) at T=298.15~K
and p=1~atm, using the B3LYP functional\protect\cite{G98}. The
first column gives the electronic energy contribution, the second
the zero-point energy, the third the energy, the fourth the
enthalpy, and the fifth column is the Gibbs free energy, all at
298.15~K.}

\label{T_2}

\begin{tabular}{lddddd}& ${\Delta}E_e$& ${\Delta}E{_0}$&
${\Delta}E_{298}^{(0)}$&
${\Delta}H_{298}^{(0)}$&${\Delta}G_{298}^{(0)}$  \\ \hline
Li$^+$+H$_2$O $\rightarrow$ Li(H$_2$O)$^{+}$ &-35.9 &-33.8 &-33.4
&-33.4 &-36.0\\ Li(H$_2$O)$^+$+H$_2$O $\rightarrow$
Li(H$_2$O)$_2$$^{+}$ &-32.2 &-30.1 &-29.9 &-30.4 &-22.5\\
Li(H$_2$O)$_{2}^+$+H$_2$O $\rightarrow$ Li(H$_2$O)$_3$$^{+}$
&-23.8 &-21.7 &-21.5 &-22.1 &-13.5\\ Li(H$_2$O)$_{3}^+$+H$_2$O
$\rightarrow$ Li(H$_2$O)$_4$$^{+}$ &-17.1 &-15.5 &-15.0 &-15.5
&-7.5\\ Li(H$_2$O)$_{4}^+$+H$_2$O $\rightarrow$
Li(H$_2$O)$_5$$^{+}$ &-9.2 &-7.3 &-7.1 &-7.7 &2.1\\
Li(H$_2$O)$_{5}^+$+H$_2$O $\rightarrow$ Li(H$_2$O)$_6$$^{+}$
&-9.8 &-6.9 &-7.3 &-7.9 &4.0\\\hline \end{tabular} \end{table}

The quantity -$\ln p_0$ of Eq.~(\ref{eqostate}) gives the free energy
in units $\beta^{-1}$ required to open this defined proximal volume in
the solvent in order that the construction of the complexes of
Eq.~(\ref{reaction}) can be pursued.  This quantity is an object of
research in the theory of liquids \cite{pratt:99} in its own right.
For specific applications an approximate form must be used.  In the
present example of the hydration of the Li$^+$ ion, this packing
contribution is expected to be of a secondary size, provided the
thermodynamic state is not varied too broadly.

Finally, the quantity $q_{Li^+}$ of Eq.~(\ref{eqostate}) is the
canonical partition function for a single Li$^+$ in a volume V at
temperature T\cite{mcq}.  The initial term of Eq.~(\ref{eqostate}) is
recognized as the ideal (no interactions) contribution to
$\mu_{Li^+}$. Thus, the second and third terms there express the
interactions of the Li$^+$ ion with the solvent and we will refer to
these contributions as $\beta\Delta\mu_{Li^+}$.

As a point of reference, for the present example of the Li$^+$ at
infinite dilution in water under the normal conditions of T=298.15~K
and $\rho_W$=1~g/cm$^3$, we calculate the quasi-chemical contributions
(the last term) to Eq.~(\ref{eqostate}) to be -128~kcal/mole using
R$_{Li^+}$=2.0\AA.  The experimental values for the total
$\Delta\mu_{Li^+}$ are -114~kcal/mol\cite{marcus} and
-125~kcal/mol\cite{Friedman:73}.\footnote{The value found in
\protect{\cite{Friedman:73}} was lowered by RT $ \ln $
[RT/(1~atm~liter/mol) ]  = 1.9~kcal/mol to convert to the standard
state adopted here.} The contributions of repulsive interactions will
raise that calculated value toward the experimental whole, but it is
unlikely to raise it enough to exceed the highest of these values.
Though the agreement is satisfactory, it seems likely that the
calculated result is too low, meaning that the Li$^+$ ion is less well
bound to liquid water than the current calculation predicts.
Additionally, the clusters have been treated in the harmonic
approximation in obtaining this theoretical value\cite{rempe:99}.

\begin{table}
\caption{Construction of aqueous solution thermochemical results
(kcal/mole) at T=298.15~K. The first column gives the gas-phase free
energy from Table~2 for the sequential hydration
reactions, the second reports the reaction net free energy after
adjustment for the actual concentration of liquid water, and the third
column gives the interaction part of the chemical potential change,
$\Delta\mu^*=\mu^*_{Li(H_2O)_n{}^+}-\mu^*_{H_2O}-\mu^*_{Li(H_2O)_{n-1}{}^+}$, 
for
n=$1,2,\ldots 6$, approximated with the dielectric model.  The fourth
column is the total free energy change for the indicated reaction in
aqueous solution, the sum of the second and third columns. To
construct the results of Fig.~2, use
$\mu^*_{Li^+}\approx$ -82.0~kcal/mol, the Born model
result for R$_{Li^+}$=2.0~\AA. }

\label{T_3}

\begin{tabular}{lddddd}&
${\Delta}G_{298}^{(0)}$&${\Delta}G_{298}$& ${\Delta}\mu^*$&
${\Delta}G_{298}$(aq) \\ \hline Li$^+$+H$_2$O $\rightarrow$
Li(H$_2$O)$^{+}$ &-36.0 &-40.3&17.3 &-23.0 \\ Li(H$_2$O)$^+$+H$_2$O
$\rightarrow$ Li(H$_2$O)$_2$$^{+}$ &-22.5&-26.8&13.3&-13.5\\
Li(H$_2$O)$_{2}^+$+H$_2$O $\rightarrow$ Li(H$_2$O)$_3$$^{+}$
&-13.5&-17.8&11.3&-6.5\\ Li(H$_2$O)$_{3}^+$+H$_2$O $\rightarrow$
Li(H$_2$O)$_4$$^{+}$ &-7.5&-11.7&8.3&-3.4\\ Li(H$_2$O)$_{4}^+$+H$_2$O
$\rightarrow$ Li(H$_2$O)$_5$$^{+}$ &2.1&-2.2& 10.3&8.1\\
Li(H$_2$O)$_{5}^+$+H$_2$O $\rightarrow$ Li(H$_2$O)$_6$$^{+}$
&4.0&-0.3& 10.3&10.0\\ \hline \end{tabular} \end{table}

\subsection*{A Thermodynamic Model} 
\begin{figure} [t!]
\centerline{\epsfig{file=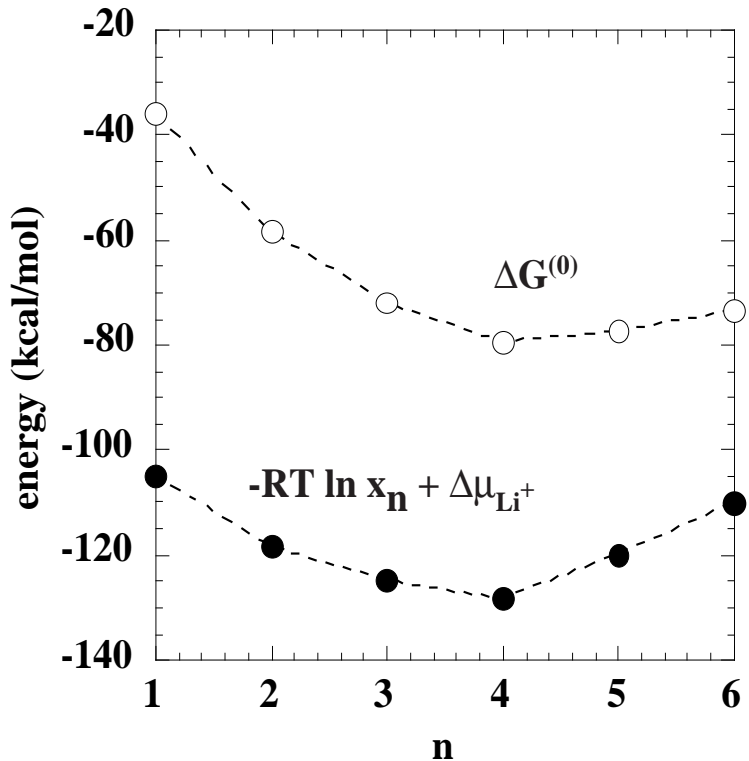,height=4.0in,width=4.0in}}
\caption{Free energies for Li$^+$(aq) ion hydration  in
liquid water plotted as a function of the number, $n$, of inner shell
water neighbors, T=298.15~K; see Table~\protect\ref{T_3}. The results
marked $\Delta$G$^{(0)}$ are the free energies predicted by the
electronic structure calculations for the reaction Li$^+$ + $n$H$_2$O
=
 Li(H$_2$O)$_n{}^+$ under standard ideal
conditions, including p = 1~atm.  The minimum value is at $n$=4.  The
solid circles labeled -RT$\,\ln \,x_n\, +\Delta \mu_{Li^+}$ are the
final, net values that include the $\rho_{H_2O}{}^n$ factors of
Eq.~(\protect\ref{eqostate}) adjusting the water molecule
concentration from the ideal case to the actual concentration of
liquid water, and the extra-cluster interaction contributions of
Eq.~(\protect\ref{ktilde}).  The final results predict that the
$n$=4 inner sphere structure is most probable in liquid water under
normal conditions.}
\label{xn}
\end{figure}

One physical view of the
quasi-chemical approach is the following: Li$^+$ ions in aqueous
solution can, in principal, be polled for the number $n$ of water
molecule ligands each possesses.  We could determine the
concentrations of Li$^+$ ions having precisely $n$ water molecule
ligands.  In fact, the $n^{th}$ term in the sum of
Eq.~(\ref{eqostate}) is proportional to those concentrations. 
Denoting the mole fraction of Li$^+$ ions with $n$ water molecule
ligands by $x_n$, successive terms there can be expressed as $x_n
\exp(-\beta\Delta\mu_{Li^+})$\cite{Pratt:98}. Those mole fractions,
however, are not independent thermodynamic variables but are governed
by the principles of chemical equilibrium.  In our example of
Li$^+$(aq), the quasi-chemical theory predicts the mole fractions of
Li$^+$ ions with $n$ water molecule ligands, as is shown in
Fig.~\ref{xn}.

These ideas can be the basis of an unrefined but almost
thermodynamical derivation of the pattern in
Eq.~(\ref{eqostate}).\footnote{This argument is conceptually simplest
and most simply expressed for the case of a chemically distinct solute
at infinite dilution.  So we limit this presentation to that case.} To
develop this `derivation', we first make explicit that we consider the
concentrations of the Li(H$_2$O)$_n^+$ complexes of different size $n$
despite the fact that it is the concentration of Li$^+$ ions that can
be experimentally manipulated.  Thus, we introduce a new set of
composition variables that refer to the ion complexes,

\begin{eqnarray} N_{Li^+}&=&\sum\limits_{n\ge 0} {{\tilde N}
_{Li\left( {H_2O} \right)_n^+}},  \nonumber \\
 N_{H_2O}&=&\tilde N_{H_2O}+\sum\limits_{n\ge 0}
{n\tilde N_{Li\left( {H_2O} \right)_n^+}}. \label{cmass.w}
\end{eqnarray}

\noindent
Here the total number of each species in solution, $N_{Li^+}$ and
$N_{H_2O}$, is expressed in terms of ${\tilde N}_{Li\left( {H_2O}
\right)_n^+}$,  the numbers of complexes of different sizes $n$,  and
${\tilde N}_{H_2O}$,  the number of water molecules not complexed to
an ion.

For the problem initially addressed in terms of variables  
$N_{H_2O}$ and $N_{Li^+}$, these relations provide a translation to
express the problem in terms of an alternative set of variables such
as $\tilde N_{H_2O}$ and $\tilde N_{Li\left( {H_2O} \right)_n^+}$. In
the formal definition of the Gibbs free energy $G$, for example, with
these changes of variables the multiplier of $\tilde N_{Li\left(
{H_2O} \right)_n^+}$ becomes $\left( \mu _{Li^+}+n\mu _{H_2O}
\right)$. If the molecule number $\tilde N_{Li\left( {H_2O}
\right)_n^+}$ could be varied independently of other such composition
variables, we would identify this multiplier as the chemical potential
conjugate to $\tilde N_{Li\left( {H_2O} \right)_n^+}$. In contrast,
for the problem initially formally addressed in terms of variables
like $\tilde N_{H_2O}$ and $\tilde N_{Li\left( {H_2O}\right)_n^+}$, as
we do here, then the relative concentrations, $x_n = {{\tilde N}
_{Li\left( {H_2O} \right)_n^+}}/N _{Li^+}$, provide a translation to
the problem expressed in terms of the standard variables such as
$N_{H_2O}$ and $N_{Li^+}$. The coefficient of $N _{Li^+}$ in the Gibbs
free energy expression is

\begin{eqnarray} 
\sum\limits_{n\ge 0} {x_n\left( {\tilde \mu _{Li\left(
{H_2O} \right)_n^+}-n{\tilde\mu} _{H_2O}} \right)} \equiv \mu_{Li^+} ,
\label{mulithium} \end{eqnarray}

\noindent the thermodynamic chemical potential conjugate to $N
_{Li^+}$.  The differences within the parenthesis depend on relative
concentrations of complexes of different sizes. These relative
concentrations are established by nontrivial conditions of
equilibrium.

To show the consequences of those conditions of equilibrium, we adopt
the standard form of the chemical potential,~\footnote{In the limit of
low density, or if the interactions are neglected, then $q_\alpha$ is
the canonical partition function for one molecule (or complex) of type
$\alpha$ in volume V at temperature T.  With a more general
understanding of the divisor of $\rho_\alpha$, the form asserted for
the ideal case is more generally valid.  It is with the molecular
identification of these factors that the present derivation goes
beyond a purely thermodynamic analysis.  We will return to this point
below.}

\begin{eqnarray}
\beta\mu_\sigma=\ln\left[\rho_\sigma V/q_\sigma\right],
\label{igas}
\end{eqnarray}

\noindent so that

\begin{eqnarray}
\beta\mu_{Li^+} = \sum\limits_{n\ge 0}
{x_n\ln\left[{\rho_{Li^+}x_n(q_{H_2O}/V){}^n\over
\rho_{H_2O}{}^n(q_{Li\left( {H_2O}
\right)_n^+}/V)} \right]} .
\end{eqnarray}

\noindent The final essential step in this thermodynamic model is the
quasi-chemical step according to which the relative concentrations of
the ion complexes of different sizes are expressed as a function of
the chemical equilibrium constants,

\begin{eqnarray}
x_n={{K_{n}^{(0)}\rho _{H_2O}{}^n} \over {\sum\limits_{m\ge 0}
{K_{m}^{(0)}\rho _{H_2O}{}^m}}} .
\label{cluster-var}
\end{eqnarray}

\noindent 
Consistent with the notation above, chemical equilibrium requires that

\begin{eqnarray}
K_{n}^{(0)}={{\left( {q_{Li(H_2O)_n^+}/V} \right)} \over {\left(
{q_{Li(H_2O)_{n=0}^+}/V} \right)\left( {q_{H_2O}/V}
\right)^n}} .
\label{Kn}
\end{eqnarray}

\noindent The K$_{n}^{(0)}$  are
a function of the temperature T as the only thermodynamic variable.
When the results above are collected, we remember that the x$_n$ are
non-negative weights that sum to one and we notice that $q_{Li^+}=
q_{Li(H_2O)_{n=0}^+}$ so that we can separate the correct ideal
contribution to $\mu_{Li^+}$.  Thus the chemical potential of the
lithium ion separates into ideal and non-ideal contributions, with the
latter contribution written in terms of the equilibrium ratios
K$_{n}^{(0)}$ that appeared in Eq.~(\ref{eqostate}),

\begin{eqnarray}
\beta\mu_{Li^+} = \ln\left[{\rho_{Li^+}\over (q_{Li^+}/V)} \right]
-\ln\left[\sum\limits_{n\ge 0} {K_{n}^{(0)}\rho _{H_2O}{}^n} \right]
.
\label{tqca}
\end{eqnarray}

\noindent Within  the simplified context of this derivation, this is 
the result that was sought.  

The previous equation should be compared with Eq.~(\ref{eqostate}).
Note that it doesn't address the initial packing contribution -$\ln
p_0$, or the other extra-cluster interaction contributions
incorporated into ${\tilde K}_n$ of Eq.~(\ref{eqostate}).  Furthermore,
it finesses the {\em recognition} issue of defining cluster structures
for counting.  Nevertheless, we can note that the quasi-chemical step,
Eq.~(\ref{cluster-var}), would arise in a pedagogical context by
making the free energy stationary with respect to variations in the
progress of chemical reactions. Because of this variational aspect
these approaches are sometimes referred to as {\em cluster-variation}
theories\cite{brush}.

\subsection*{Variations of Hydration Free Energies} Predictions of
{\em variations} of hydration free energy changes with temperature,
pressure, composition, and solute geometry are the foremost practical
weaknesses of dielectric continuum models for hydration free energies.
An important facet of this issue is that the radii parameters in the
dielectric models, which are initially established fully empirically,
depend on solution thermodynamic conditions such as temperature,
pressure, and composition, and on solute
conformation\cite{Pratt:94a,Tawa:94,Tawa:95,Pratt:97,Hummer:98}.  The
Li$^+$(aq) case gives direct example of this difficulty: the partial
molar volume of the Li$^+$(aq) is {\em negative},
-6.4~cm$^3$/mol\cite{marcus}.  The magnitude of this important
thermodynamic quantity is not explicable simply on the basis of the
Born model\cite{marcus} and the {\em sign} cannot be rationalized
either if any reasonable description of excluded volume effects is
included.  At this point, considerable further evolution of dielectric
models typically occurs with additional empirical
parameters\cite{marcus,Floriano:98}. In contrast, an important feature
of the present approach is that explicit contributions to the
temperature and density dependences of the chemical potential are
readily available.  This facilitates investigation of thermodynamic
derivatives.

\subsubsection*{Pressure Variation of Hydration Free Energies} The
variation of the hydration free energies with pressure is the partial
molar volume and gives direct information on hydration structure.
Thus, it is particularly valuable in learning about the hydration
process. Recalling that the chemical potential expression used in the
current approach is written explicitly in Eq.~(\ref{eqostate}), the
partial molar volume $\left.  {\left(
\partial V /
\partial N_{Li^+} \right)}
\right|_{\beta,p,N_{H_2O}}$ is

\begin{eqnarray} v_{Li^+} & \equiv & \left. {\left( {{{\partial
\mu _{Li^+}} \over {\partial p}}} \right)}
\right|_{\beta,N_{H_2O,}N_{Li^+}} \nonumber \\ & = & {1 \over
\rho _{Li^+}} \left. \left( {\partial \rho _{Li^+}\over \partial
\beta p} \right) \right|_{\beta,N_{H_2O},N_{Li^+}}  + \left( {
\partial \beta \Delta \mu _{Li^+} \over \partial \rho _{H_2O}}
\right)_{\beta,\rho _{Li^+}} \left. \left( {\partial \rho _{H_2O}
\over \partial \beta p} \right) \right|_{\beta,N_{H_2O},N_{Li^+}}
 . \end{eqnarray}

\noindent Here we are interested exclusively in the conditions of
infinite dilution of the aqueous solution so that

\begin{eqnarray} \lim_{\rho _{Li^+}\rightarrow 0}v_{Li^+}=(\kappa
_T/\beta)\left[ {1+\rho _{H_2O}\left( {{{\partial \beta \Delta
\mu _{Li^+}} \over {\partial \rho _{H_2O}}}} \right)_\beta}
\right], \label{pmv.01} \end{eqnarray}

\noindent where $\kappa_T$ = $(-1/V)\left( \partial V / \partial
p \right)_T$ is the isothermal coefficient of bulk compressibility of
the pure solvent.  Upon differentiation, the first two terms of
Eq.~(\ref{eqostate}) produce the partial molar volume of a hard
inert sphere of the right size at infinite dilution; call it $ v_0$.
As indicated above this is a topic of separate study\cite{palma:93}.
Thus,

\begin{eqnarray} \lim_{\rho _{Li^+}\rightarrow 0} \, v_{Li^+} & = &
v_0 - (\rho_{H_2O}RT\kappa_T) \left( {\partial \over \partial
\rho _{H_2O}}\right)_\beta \ln \left[ \sum\limits_{n \ge 0}
{\tilde K_{n}}\,\rho _{H_2O}{}^n \right] . \label{pmv.02}
\end{eqnarray}

\noindent If the initial approximation ${\tilde K_{n}}\approx
K_{n}^{(0)}$ is retained, or more generally if the extra-complex, or
implicit solvent parts are only weakly density dependent, then the
derivative indicated in Eq.~(\ref{pmv.02}) can be evaluated
explicitly.  Within this approximate view, the formula in
Eq.~(\ref{pmv.02}) can be given a direct interpretation. We can parse
the final term in Eq.~(\ref{pmv.02}) as

\begin{eqnarray}
\Delta v = (\kappa_T\rho_{H_2O}RT)\; v_{H_2O}\; {\overline n}
\label{simple}
\end{eqnarray}
where $v_{H_2O} = 1/\rho_{H_2O}$ is the partial molar volume of
the pure solvent (water) and
\begin{eqnarray} {\overline n} =  \left( {\partial \over \partial
\ln\rho_{H_2O}}\right)_\beta \ln \left[ \sum\limits_{n  \ge 0} 
{\tilde K_{n}}\,\rho _{H_2O}{}^n \right] = \sum\limits_{n \ge 0}n x_n
. \label{pmv.03}
\end{eqnarray}

\noindent The leading factor $\rho_{H_2O}RT\kappa_T$ appropriately
converts a density change to a pressure change. The following factors
assess the volume change per ligand and the number of ligands.  This
interpretation doesn't work in this specific sense if the density
dependence of the coefficients in Eq.~(\ref{pmv.03}) is significant.
(It is interesting to note, however, the more general interpretation
given by References~\cite{Matubayasi:94,Matubayasi:96} to the partial
molar quantities.)  

\begin{figure}
\hspace{1.0in}
\centerline{\epsfig{file=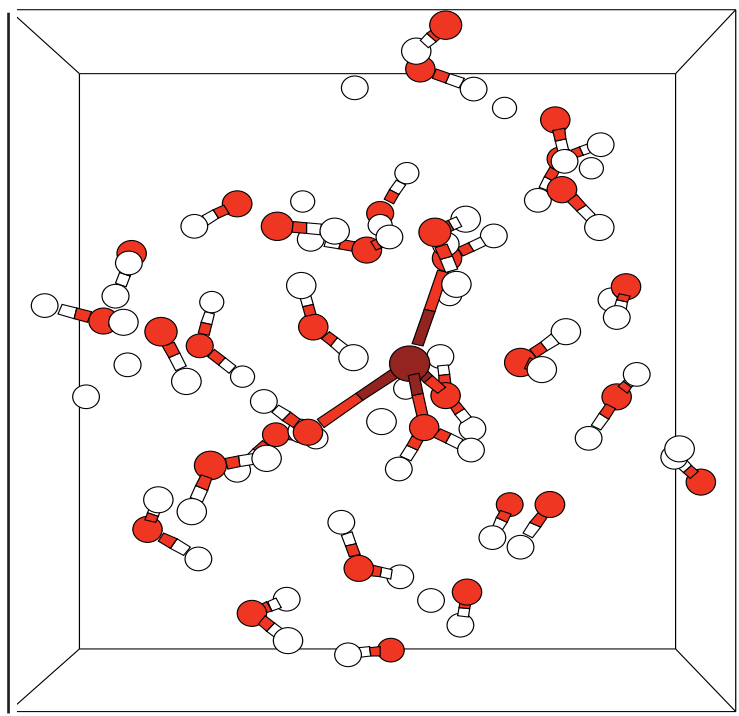,height=4.7in,width=4.38in}}
\caption{A structure from `ab initio' molecular dynamics calculations
based upon a gradient-corrected electron density functional
description of the interatomic forces, after
Ref.~\protect\cite{rempe:99}.  Initially a hexa-coordinate inner
sphere structure, rigidly constrained, was equilibrated with 26
additional water molecules by Monte Carlo calculations using classical
model force fields.  The structure shown, with $n=4$ water ligands,
resulted 125~fs later and remained the dominant inner shell complex
for the duration of the simulation.  The bonds drawn in identify water
oxygen atoms within 2.65~\AA\ of the Li$^+$ ion.} \label{vasp}
\end{figure}

We find ${\overline n}$ = 4.0, as demonstrated in Fig.~\ref{xn} in the
present example of the Li$^+$ at infinite dilution in water under the
normal conditions of T=298.15~K and $\rho_W$=1~g/cm$^3$. As depicted
in Fig.~\ref{vasp}, the value ${\overline n}$ = 4.0 has been
corroborated by `ab initio' molecular dynamics
calculations\cite{rempe:99}.  Combined with the other factors,
${\overline n}$ = 4.0 produces a contribution of -4.9~cm$^3$/mole to
the partial molar volume of Eq.~(\ref{simple}).  The simplest idea for
estimation of the non-electrostatic contributions is to use the
experimental value for the partial molar volume of He as a solute in
water, 14.8~cm$^3$/mole\cite{clever}, making the total predicted
partial molar volume of Li$^+$ $v_{Li^+}$=9.9~cm$^3$/mole.  Therefore,
this pain-staking accounting of the effects through the first
hydration shell, detailed though it is, does not satisfactorily
explain the partial molar volume of Li$^+$ in liquid water.  Since
Li$^+$(aq) is known to have a strongly structured second hydration
shell\cite{Heinzinger:79,Mezei:81,Impey:83,Chandrasekhar:84,%
bounds:85,Zhu:91,Romero:91,lee:94,toth:96,Koneshan:98b}, use of the
dielectric model for the interactions of the Li$^+$ with the outer
hydration shells is a concern and may account for the difference
between the calculated and experimental partial molar volumes.

We note that Eq.~(\ref{simple}) assumes that the implicit
contributions to the pressure variation are negligible.  Use of the
Born model directly without an empirical attribution of pressure
dependence to the Born radii but with the measured pressure
dependence of the solvent dielectric constant\cite{nbs} contributes
-0.3~cm$^3$/mol, a negligible magnitude here.  The density dependences
of hydration free energies for the auto-dissociation reaction $2H_2O
\rightleftharpoons H_3O^+ + OH^-$ in water were studied some time
ago\cite{Tawa:95} from the point of view of a dielectric continuum
model with the conclusion that a non-trivial density dependence was
required of the dielectric model by the experimental data.  In
contrast, the formula Eq.~(\ref{simple}) is properly insensitive to
radii parameters assigned to the solute.

\subsubsection*{Temperature Variation of Hydration Free Energies} The
temperature variation of the hydration free energy is the partial
molar entropy and, because of its interpretation as an indicator of
disorderliness, is of wide interest.  As before, we focus here on the
conditions of infinite dilution in the aqueous solution.  Again, the
first two terms of Eq.~(\ref{eqostate}) produces the partial molar
entropy of a hard inert sphere of the right size at infinite
dilution\cite{garde:96}.  We will call that contribution $s_0$.
Differentiation of the quasi-chemical contributions in
Eq.~(\ref{eqostate}) with respect to temperature at constant pressure
yields the partial molar entropy:

\begin{eqnarray} 
\lim_{\rho _{Li^+}\rightarrow 0}
s_{Li^+}
& = & 
   s_0 + \left[ { \left( {{\partial \rho_{H_2O}} \over {\partial T}} 
         \right)}_{p}
{\left( { {\partial } \over {\partial \rho _{H_2O}}}
        \right)}_{T}
 + 
{\left( { {\partial  } \over {\partial T} }
        \right)}_{\rho _{H_2O}} \right]\left( RT \ln \left[
\sum\limits_{n \ge 0} {\tilde K_{n}}\,\rho _{H_2O}{}^n 
\right]\right)
\label{pms.01} 
\end{eqnarray}

\noindent
The middle term accounts for the temperature dependence of the solvent
density, which brings in the coefficient of thermal expansion for the
pure solvent $\alpha _p$ =$(1/V)(\partial V/\partial T)_p$, and then
requires the density derivative of the quasi-chemical contributions.
That density derivative was analyzed above when we considered the
partial molar volume.  Using Eq.~(\ref{pmv.02}) produces

\begin{eqnarray} 
\lim_{\rho _{Li^+}\rightarrow 0}
s_{Li^+}
& = & 
   s_0 + {\alpha _p \over\kappa_T}( v_{Li^+} - v_0)
   + 
{\left( { {\partial  } \over {\partial T} }
        \right)}_{\rho _{H_2O}}\left( RT \ln \left[
\sum\limits_{n \ge 0} {\tilde K_{n}}\,\rho _{H_2O}{}^n 
\right]\right)  \nonumber \\
 & = & 
 s_0 + {\alpha _p \over\kappa_T}( v_{Li^+} - v_0) + R \left( {1-\beta
 \left( {{\partial \over {\partial \beta }}}
 \right)_{\rho _{H_2O}}} \right) \ln \left[
 \sum\limits_{n \ge 0} \tilde{K_{n}}\rho _{H_2O}{}^n \right].
\label{pms.02} 
\end{eqnarray}
The developments below establish that the fundamental quantities
${\tilde{K_{n}}}$ are well-defined and therefore the temperature
derivative indicated here can be investigated with well defined
procedures.  For the purposes of this specific section and example, we
will insist on the approximation ${\tilde{K_{n}}}\approx K^{(0)}_n$.
In that case, the last form of Eq.~(\ref{pms.02}) highlights
quasi-chemical contributions to standard enthalpy changes for the
reactions in Eq.~(\ref{reaction}) because
\begin{eqnarray}
\left( {{\partial \over {\partial \beta }}} \right)_{\rho
_{H_2O}} \ln \left[ \sum\limits_{n \ge 0} K^{(0)}_n \rho
_{H_2O}{}^n \right]=-\sum\limits_{n\ge 0} {x_n\Delta
H_n^{(0)}} ,
\label{enthalpy}
\end{eqnarray}
with $x_{n}$ the normalized relative concentrations of the
Li(H$_2$O)$^+_n$ complexes defined earlier in Eq.~(\ref{cluster-var})
and $\Delta H^{(0)}_n$ the enthalpies of reaction for the ion
hydration reactions listed in Table~\ref{T_2}.  With this
identification Eq.~(\ref{pms.02}) then becomes

\begin{eqnarray} 
\lim_{\rho _{Li^+}\rightarrow 0} s_{Li^+} & \approx &
 s_0 + ( v_{Li^+} - v_0){\alpha _p \over\kappa_T} + 
R\; \ln \left[ \sum\limits_{n \ge 0} {K^{(0)}_{n}}\rho _{H_2O}{}^n \right]
+ {1\over T} \sum\limits_{n \ge 0} x_{n} \Delta H^{(0)}_{n}.
\label{pms.04} 
\end{eqnarray}
The term associated with the quasi-chemical contributions to the
partial molar volume here contributes -0.5~cal~K$^{-1}$~mol$^{-1}$ to
the partial molar entropy. The subsequent terms can be evaluated from
Tables~\ref{T_2} and \ref{T_3}, yielding
-16.1~cal~K$^{-1}$~mol$^{-1}$. Again using the experimental value for
He as a solute in water to estimate the non-electrostatic contribution
produces $s_0$ =
-9.3~cal~K$^{-1}$~mol$^{-1}$\cite{clever}.\footnote{The value found in
\protect\cite{clever} was augmented by R~$\ln$[$\rho_{H_2O}$RT] =
14.3~cal~K$^{-1}$~mol$^{-1}$, with $\rho_{H_2O}$ = 1~g/cm$^3$, to
transform to the standard state adopted here.} The combined value for
the partial molar entropy of lithium ion is
-25.9~cal~K$^{-1}$~mol$^{-1}$ compared to experimental values of
-38.5~cal~K$^{-1}$~mol$^{-1}$\cite{marcus} and
-40.1~cal~K$^{-1}$~mol$^{-1}$\cite{Friedman:73}.\footnote{The value
found in \protect\cite{Friedman:73} was adjusted by
R~$\ln$[RT/(1~atm~liter/mol)] = -6.4~cal~K$^{-1}$~mol$^{-1}$ to convert
to the standard state adopted here.} This calculated partial molar
entropy is less negative than the experimental value.  Our neglect of
the outer hydration shells may explain that difference.  In this
context, however, the comparison given by Friedman and
Krishnan\cite{Friedman:73} of the hydration entropies of K$^+$ +
Cl$^-$ and Ar + Ar is particularly relevant: the hydrophobic
contributions to these {\em negative} entropies of hydration can be
the principal contributions.  Here that hydrophobic contribution has
been estimated crudely using the He results as a model.

It is striking that this careful account of the inner hydration shell
structure, with the complete neglect of effects of the more distant
medium and thus neglect of dielectric properties, produces such a
substantial part of the experimental partial molar entropy.  It is
similarly striking that the quasi-chemical contribution to the partial
molar volume, and specifically ${\overline n}$, is insensitive to the
dielectric model utilized.  This suggests the hypothesis that the
popular dielectric continuum models can be satisfactory with some
empiricism for the hydration free energies but such dielectric effects
contribute only secondarily to variations in hydration free energies.

\section*{Quasi-Chemical Theory}
This section gives a detailed derivation of the quasi-chemical pattern
of calculation for statistical thermodynamics.  This derivation
discusses a variety of details in order to provide additional
perspective on the mechanisms of the theory. The construction of the
derivation depends on two basic ingredients: the potential
distribution theorem and a clustering analysis. These developments
culminate in a generalization of the thermodynamic model above and a
{\em quasi-chemical rule}, Eq.~(\ref{gqca}), that is then given a
succinct direct proof.

\subsection*{Potential Distribution Theorem} 
According to the potential distribution theorem\cite{Widom:82}

\begin{eqnarray}
\rho _\sigma =\left\langle {e^{-\beta \Delta U}} \right\rangle
_0z_\sigma \left( {q_\sigma /V} \right),
\label{widom}
\end{eqnarray}

\noindent where $\rho_\sigma$ is the density of molecules of type
$\sigma$ (the `solute' under consideration), $z_\sigma =
\exp(\beta\mu_\sigma)$ is the absolute activity of that species,
q$_\sigma$ is the single molecule partition function for that species,
and V is the volume.  The indicated average is the usual test particle
calculation: the average over the thermal motion of the bath of the
interaction potential energy between the test particle and the bath.
A virtue of this result is that the indicated average is similar to a
partition function but local in character.  Thus all the concepts
relevant to evaluation of a partition function are relevant here too,
but they typically work out more simply because of the local character
of this formulation.

The potential distribution formula, Eq.~(\ref{widom}), can be
directly generalized to treat conformational problems:\cite{ecc}

\begin{eqnarray}
\rho _\sigma ({\bf 1}) = s_\sigma ^{(0)}({\bf 1})\left\langle 
{e^{-\beta \Delta U}} \right\rangle _{0,{\bf 1}}z_\sigma q_\sigma 
\label{conformation} 
\end{eqnarray}

\noindent where  $s_\sigma^{(0)}({\bf 1})$ is the isolated molecule
distribution function over single molecule conformational space,
including translations and rotations.\footnote{It is worth emphasizing
in the context of dielectric models that $\left\langle {e^{-\beta
\Delta U}} \right\rangle _{0,{\bf 1}}$ involves the charge
distribution of the isolated molecule, not the `self-consistent'
charge distribution that has relaxed to accomodate interactions with
the solution external to the solute.} This function is normalized so
that

\begin{eqnarray}
\int {s_\sigma ^{(0)}({\bf 1})d{\bf 1}}=1 .
\end{eqnarray}

\noindent The additional subscript on the 
$\left< \ldots \right>_{0,{\bf 1}}$ brackets in Eq.~(\ref{conformation}) 
indicate that the average is obtained for the
solute in conformation {\bf 1}.  This generalization suggests a
notational simplification according to which Eq.~(\ref{widom}) can be
expressed as

\begin{eqnarray}
\rho _\sigma = \left\langle {\left\langle {e^{-\beta \Delta U}}
\right\rangle } \right\rangle _0z_\sigma \left( {q_\sigma /V} \right) .
\label{double}
\end{eqnarray}

\noindent Now the double brackets $\left<\left<\ldots\right>\right>_0$
indicate the average over the thermal motion of the solute {\em and}
the solvent under the conditions of no interaction between them, and
the averaged quantity is the Boltzmann factor of those interactions.
%Rearrangement of this equation yields a ratio of activities equal to the average,
%where the ratio consists of 
%the activity of the ideal solute, $\rho_\sigma/{(q_\sigma/V)}$,  divided by
The average indicated here is the ratio of
%the activity of the ideal solute, $\rho_\sigma/{(q_\sigma/V)}$,  divided by
the activity of an isolated solute, $\rho_\sigma V/{q_\sigma}$,
divided by the absolute activity, $z_\sigma$, of the actual solute.
In the developments that follow we will focus on the resulting feature

\begin{eqnarray}
\left\langle \left\langle e^{-\beta \Delta U}
\right\rangle  \right\rangle _0 \equiv e^{-\beta
\Delta \mu _\sigma},
\label{average}
\end{eqnarray}

\noindent  performing series expansions and otherwise analyzing this 
average.

\subsection*{Clustering Analysis}  The second basic ingredient to
deriving the quasi-chemical pattern is a counting device used to sort
configurations according to proximity.  It can be motivated by
focusing again on the averaged quantity in Eq.~(\ref{average}).  If
$\Delta$U were pairwise decomposible we would write

\begin{eqnarray}
e^{-\beta \Delta U}=\prod\limits_j {(1+f_{\sigma j})} ,
\end{eqnarray}

\noindent  where f$_{\sigma j}$ is the Mayer f-(cluster)-function
describing the interactions between the $\sigma$ solute and the
solvent molecule $j$\cite{uhlenbeck}. Series expansions then proceed
in a direct and simple way for this case.  The clustering features of
these expansions are valuable and we can preserve them in cases that
don't present pairwise decomposable $\Delta$U by writing

\begin{eqnarray}
1=\prod\limits_j {(1+b_{\sigma j}+f_{\sigma j})} .
\label{clustering}
\end{eqnarray}

\noindent   b$_{\sigma j}$ is one (1) in a geometrically defined
$\sigma j$-bonding region and zero (0) otherwise; b$_{\sigma j}$ is an
indicator function\cite{vankampen:indicator} or an `inclusion
counter.'  f$_{\sigma j}$ is then defined by

\begin{eqnarray}
f_{\sigma j}=-b_{\sigma j} .
\label{Mayerf}
\end{eqnarray}

\noindent   With this setup the f$_{\sigma j}$
is entirely analogous to a Mayer f-function for a hard object and can
play the same role of monitoring the description of packing effects in
liquids.  
%In contrast to b$_{\sigma j}$, which is an inclusion counter
%used to define the bonding region, f$_{\sigma j}$ is an exclusion counter
%used to define the non-bonding region.
Fig.~\ref{fig0} shows the natural definition of the bonding region
for the application to spherical atomic ions: b$_{\sigma j}$ is one
(1) inside the sphere that includes the inner shell of water molecules
whereas f$_{\sigma j}$ is negative one (-1) there.  Both b$_{\sigma
j}$ and f$_{\sigma j}$ are zero (0) outside of this bonding region.

The strategy for our derivation will be to insert this `resolution of
1,' Eq.~(\ref{clustering}), within the averaging brackets of the
potential distribution theorem, then expand and order the
contributions according to the number of factors of b$_{\sigma j}$
that appear. We emphasize that physical interactions are not addressed
here and that the `hard sphere interactions' appear for counting
purposes only.

\subsubsection*{Low Order Contributions} Let's note some 
of the properties of this expansion and the terms that result.
Consider initially the term that is zeroth order, no factors of
b$_{\sigma j}$ appearing.  That term is

\begin{eqnarray}
0^{th}\;order\;in\;b_{\sigma j}: \prod\limits_j {(1+f_{\sigma j})} .
\label{zeroth}
\end{eqnarray}

\noindent  This would be the interaction potential energy for the bath 
with a hard object that excludes the bath from any bonding region
because then f$_{\sigma j}$ =-1 and the statistical weight
of Eq.~(\ref{zeroth}) would be zero. %then.  
Next

\begin{eqnarray}
1^{st}\;order\;in\;b_{\sigma j}: Nb_{\sigma 1}\prod\limits_{j\ge 2}
{(1+f_{\sigma j})} .
\label{first}
\end{eqnarray}

\noindent  There is just one bath/solvent molecule in the bonding 
region and N possibilities for that specific molecule due to the
existence of N solvent molecules in the system.  All other
molecules are excluded from the bonding region.  Next

\begin{eqnarray}
2^{nd}\;order\;in\;b_{\sigma j}: \left( {{{N\left( {N-1} \right)}
\over 2}} \right)b_{\sigma 1}b_{\sigma 2}\prod\limits_{j\ge 3}
{(1+f_{\sigma j})} .
\label{second}
\end{eqnarray}

\noindent  Now there are two bath/solvent molecules in the bonding 
region, that pair could have been chosen in N(N-1)/2 ways and all
other molecules excluded.

\subsubsection*{General Term}The pattern of these contributions to the
cluster expansion is obvious.  If we take the general formula for the
$n^{th}$ order expansion term and now include the Boltzmann factor for
the full system of solute and solvent molecules, then we can
illustrate the equivalence of two different views of the total system.
In one view the system is divided into one solute and the set of $N$
solvent molecules compared to an alternative view in which the solute
might be a cluster.  In the cluster view, the solute is surrounded by
$n$ solvent molecules in the bonding region with the remaining $N-n$
solvent molecules occupying the external region. Including the
interactions, the general term takes the form

\begin{eqnarray}
n^{th}\;& order& \;in\;b_{\sigma j}: s_\sigma ^{(0)}({\bf 1})
{e^{-\beta (U(N)+\Delta U)}\;\left( {\matrix{N\cr n\cr }}
\right)\prod\limits_{i=1}^n {b_{\sigma n}}\prod\limits_{j\ne \{1,\ldots,n\}}
{(1+f_{\sigma j})}} \nonumber \\ & = & \left( {{{n!q_{\sigma W_n}}
\over {q_\sigma}}} \right)s_{\sigma W_n}{}^{(0)}({\bf 1+n}){e^{-\beta
(U(N-n)+\Delta U)}\;\left( {\matrix{N\cr n\cr }}
\right)\prod\limits_{j\ne \{1,\ldots,n\}}
{(1+f_{\sigma j})}}.
\label{nth}
\end{eqnarray}

\noindent  
Although the expression on each side of this equation contains the
full Boltzmann factor for the complete N+1 molecule system, the
$\Delta U$ on each side differs. The final $\Delta U$ is the
difference between the potential energy of the N+1 molecule system and
the energies of the separate N-n and (1+n) systems.  Thus the final
$\Delta U$ expresses the interaction energy between the separate
cluster and external solvent systems. None of the energies considered
here need be pairwise decomposable. $q_{\sigma W_n}$ is the canonical
partition function of a cluster of n solvent molecules with the
$\sigma$ (solute) molecule in a volume V at temperature T.  In the
Li$^+$(aq) example, the Li(H$_2$O)$_n{}^+$ are the clusters and
$q_{Li\left( {H_2O}\right)_n^+}$ are the cluster partition functions.
Similarly, $s_{\sigma W_n}{}^{(0)}({\bf 1+n})$ is the canonical
configurational distribution for the complex at temperature
T.\footnote{A more technical observation: In the last member of
Eq.~(\protect\ref{nth}), n!  is asserted to be the symmetry number for
the complex.  This classically reflects the molecule exchange symmetry
of the ligand molecules.  This is correct if the ligands are identical
to one another and different from the solute that serves as a nucleus
of the cluster.  It is also correct if the ligands and nucleus are the
same species because of the structure of the bonding Boltzmann factors
requires the cluster to be of star type and the nucleus is thus
distinguishable from the ligands on that basis. Finally, the same
formula ultimately results also if the ligands are not all of the same
species. The symmetry numbers (particle exchange symmetries) of the
individual molecules aren't involved.}

It is convenient and natural to express the averaging required by the
basic potential distribution formula of Eq.~(\ref{conformation}) as a
product of averages for the two separate systems.  Since the Boltzmann
factors for the separate systems are identified in Eq.~(\ref{nth}), we
can do this merely by factoring the correct normalizing denominators
into the proper places:

\begin{eqnarray}
n^{th}\;order \; contribution \; to \left\langle {\left\langle
{e^{-\beta \Delta U}} \right\rangle } \right\rangle : \nonumber
\end{eqnarray}

\begin{eqnarray}
\left( {{{n!q_{\sigma W_n}} \over {q_\sigma}}} \right)
 \left\langle \left\langle {e^{-\beta \Delta U}\;\left( {\matrix{N\cr
n\cr }} \right) \prod\limits_{j\ne \{1,\ldots,n\}} {(1+f_{\sigma j})}}
\right\rangle _{0} \right\rangle _n \left( {Q(N-n)(N-n)! \over
Q(N)N!} \right) \nonumber \\ =
\left\langle  \left\langle {e^{-\beta \Delta U} \prod\limits_{j\ne \{1,\ldots,n\}}
{(1+f_{\sigma j})}} \right\rangle _{0} \right\rangle _n z^n{{{q_{\sigma W_n}} / {q_\sigma}}} ,
\label{finalnth} 
\end{eqnarray}

\noindent  noting that $z^n = Q(N-n)/Q(N)$ are absolute activity
factors for the solvent molecules.  The brackets $\left\langle
\left\langle \ldots \right\rangle _{0} \right\rangle _n$ indicate the
average with the distribution that is the product of the distributions
for the bath and for the complex with $n$ ligands in addition to the
solute. Finally, we define an additional quantity

\begin{eqnarray}
p_0 = \left\langle \prod\limits_{j\ne \{1,\ldots,n\}} {(1+f_{\sigma j})}
\right\rangle _0 .
\label{p0}
\end{eqnarray}

\noindent p$_0$ is independent of $n$ in the thermodynamic limit for 
finite $n$.  The quantity averaged is zero for any solvent
configuration for which some solvent molecule penetrates the defined
proximal volume of the solute.  In the simplest examples, such as the
Li$^+$ ion example above, this volume does not depend on solute
conformation.  In general, however, p$_0$ can depend on solute
conformation. Consider, for example, an extended solute with two or
more hydrophilic sites where that the proximal volume is defined to be
the union of spheres centered on each hydrophilic site.  In any case,
with knowledge of p$_0$ we can incorporate the exclusion factors of
Eq.~(\ref{finalnth}) into the bath distribution and write

\begin{eqnarray}
n^{th}\;order \; contribution \; to \left\langle {\left\langle
{e^{-\beta \Delta U}} \right\rangle } \right\rangle : \nonumber \\
\left\langle  p_0 \left\langle e^{-\beta \Delta U} \right\rangle ^*_{0}
\right\rangle _n z^n{{{q_{\sigma W_n}} / {q_\sigma}}} , \label{finalfinal}
\end{eqnarray}

\noindent The superscript * indicates that the bath distribution
function now rigidly excludes additional solvent molecules from the
defined volume.

Collecting these results and rewriting Eq.~(\ref{average}), finally we
get the structure of a partition function,

\begin{eqnarray}
e^{-\beta \Delta \mu _\sigma} = \sum_{n \ge 0} \left\langle  p_0
\left\langle e^{-\beta \Delta U} \right\rangle ^*_{0} \right\rangle
_n z^n{{{q_{\sigma W_n}} / {q_\sigma}}},
\label{gcresults}
\end{eqnarray}

\noindent 
as intended.  The systems described by this partition function, %are
however, do not constitute a familiar statistical thermodynamic
ensemble.  Instead the system is an open microscopic volume defined
relative to a specific molecule of type $\sigma$. Additionally, the
energies involved are energy differences.  The calculation identifies
$\beta \Delta \mu _\sigma$ as the thermodynamic potential determined
by this partition function directly.  This result is directly
comparable to pedagogical depictions of Bethe-Guggenheim approximate
treatments of order-disorder theory\cite{feynman}.

\subsection*{Quasi-Chemical Approximation and Generalization}

In a previous publication\cite{Pratt:98}, a specific quasi-chemical
approximation was identified that allowed a simplified expression of
the interaction part of the solute chemical potential in terms of the
solvent densities, rather than activities. A general way to achieve
this result is to apply Eq.~(\ref{average}) to the ligand (water)
species, and use the resulting expression to eliminate the activities
$z$ in Eq.~(\ref{gcresults}). If we assume that p$_0$ is independent
of conformation and that the resulting ratios of bath contributions
equal unity,\footnote{This second assumption would be credible for
systems with short-ranged forces away from a critical point and is the
heart of the quasi-chemical approximations developed for critical
phenomena.} then we obtain precisely that quasi-chemical approximation

\begin{eqnarray}
e^{-\beta \Delta \mu _\sigma} \approx p_0\sum_{n \ge 0} 
K_n^{(0)}\,\rho_{H_2O}{}^n .
\label{qca.01} \end{eqnarray} 

\noindent   The equilibrium ratio for the aggregation reaction where
$n$ ligand (water) molecules associate with the $\sigma$ species, as
in Eq.~(\ref{reaction}) with Li$^+$ ion, evaluated without
consideration of the effects of the more distant medium is
K$_n^{(0)}$.  That evaluates naturally to one (1) when n=0.  This is a
first opportunity to compare the results of these theoretical
considerations with the proposed form of the solute chemical potential
in Eq.~(\ref{eqostate}), where medium effects on the clusters are
included.  In fact, the example results of Eqs.~(\ref{simple}) and
(\ref{pms.04}) utilized this approximation.

For the problems of interest here, long-ranged external-cluster
interactions should be included, even though the
assumption that p$_0$ is independent of solute conformations can be
retained a while longer.  Since we are in full possession of the
formally complete quasi-chemical result, Eq.~(\ref{gcresults}), we can
restore the correct factors to the approximate result in
Eq.~(\ref{qca.01}) by defining\cite{Hummer:98}

\begin{eqnarray}
\tilde K_n=K_n^{(0)}\left[ {{{\left\langle {\left\langle {e^{-\beta
\Delta U}} \right\rangle _0^\ast} \right\rangle _n} \over {\left\langle
{e^{-\beta \Delta U}} \right\rangle _0{}^n}}} \right] .
\label{ktilde}
\end{eqnarray}

\noindent Here the factor in the numerator still refers to cluster and
external solvent interactions while the denominator factors refer to
the ligand (water) and external solvent interactions. The generalized
expression for the solute chemical potential becomes

\begin{eqnarray}
e^{-\beta \Delta \mu _\sigma }=p_0\sum\limits_{n \ge 0 } 
{\tilde K_n}\,\rho _{H_2O}{}^n .
\label{qca.02} 
\end{eqnarray} 

The bracketed ratio in Eq.~(\ref{ktilde}) should be amenable to
physical approximation because it should be insensitive to details of
intermolecular interactions at short-range.  One possible way to
estimate the value of the ratio is with the common dielectric
continuum model\cite{Hummer:98}. Using this model, we notice that the
factors in the denominator are similar to contributions that will be
involved in the numerator of the ratio. The contributions that aren't
balanced in this way are of two types.  The first of these are
contributions associated with the nucleus of the cluster; but that
nucleus is typically well-buried by a coating of ligands.  The second
unbalanced contribution is associated with through-solvent
interferences between different ligands.  Thus, the interactions that
aren't balanced in this desirable way are all of somewhat longer range
than the ones that are balanced. These arguments are of just the type
that might be presented to support the classic quasi-chemical
approximation of Eq.~(\ref{qca.01}).  But if a physical approximation
for external-cluster interactions is available --- as with the
dielectric models --- then the results should be insensitive to
details of the implementation, even when the quantitative contributions
aren't negligible.  In fact, this proved to be true in the lithium
example.  There it was found that increasing the radius assumed for
the Li$^+$ ion from R$_{Li^+}$=2.0\AA\ to 2.65\AA\ increased the
predicted hydration free energy by only 1\%\cite{rempe:99}.

\subsection*{Generalization of the Thermodynamic Derivation} The
thermodynamic model developed in the previous section, culminating
with Eq.~(\ref{tqca}), can be generalized to include longer ranged
correlations.  This is achieved by exploiting the potential
distribution theorem, Eq.~(\ref{widom}), through the replacement

\begin{eqnarray}
q_\sigma
\leftarrow q_\sigma \left\langle {e^{-\beta \Delta U}}
\right\rangle _0.
\label{replacement}
\end{eqnarray}

\noindent  With this replacement of the ideal solute partition
function, the mass-action relations in Eqs.~(\ref{cluster-var}) and
(\ref{Kn}) remain valid because the central thermodynamic information
in Eq.~(\ref{igas}) is correctly generalized to
$\mu_\alpha=RT\ln\left[\rho_\alpha V/q_\alpha\left\langle {e^{-\beta
\Delta U}} \right\rangle _0\right]$, as the potential distribution
theorem in Eq.~(\ref{widom}) indicates.  The equilibrium ratios K$_n$
now have the interpretation as the actual, observed ratios established
in solution

\begin{eqnarray}
K_n={\rho_{Li(H_2O)_n{}^+} \over \rho_{Li(H_2O)_{n=0}{}^+}
\rho_{H_2O}{}^n } .
\label{Kn-ob}
\end{eqnarray}

\noindent Finally we obtain the generalized expression for the
interaction part of the chemical potential,

\begin{eqnarray}
\beta\Delta\mu_{Li^+}   =  -\ln\left[ 
\left\langle {e^{-\beta
\Delta U}}\prod\limits_j {(1+f_{\sigma j})}\right\rangle_0  \right]
-\ln\left[\sum\limits_{n\ge 0} {K_n\,\rho _{H_2O}{}^n} \right] .
\label{gqca.initial} \end{eqnarray}

\noindent  We have elaborated the notation to clarify how the
remaining Widom factor refers to the solute with zero ligands, as
defined in the first logarithmic term.  The average $\langle
{e^{-\beta \Delta U}}\prod{}_j {(1+f_{\sigma j})} \rangle_0$ in
Eq.~(\ref{gqca.initial}) eliminates close solvent contacts with the
bare solute and should be well described with the aid of dielectric
continuum models.  It should be compared with the p$_0$ term of
Eq.~(\ref{qca.02}) that was constructed by consideration of `cluster
interference' issues\cite{Pratt:98}.  The difference in the structure
of this result compared to Eq.~(\ref{eqostate}) is a consequence of
arranging the expression so that the actual, observed chemical
equilibrium ratios K$_n$ appear here.

Insisting on this latter point produces the remarkable quasi-chemical
rule
\begin{eqnarray}
\beta\Delta\mu_{Li^+}  = - \ln\left[
\left\langle {e^{-\beta
\Delta U}}\prod\limits_j {(1+f_{\sigma j})}
\right\rangle_0  \right]+ \ln x_0 ,
\label{gqca} \end{eqnarray}

\noindent
where $x_0$ refers to the relative concentration of the $n=0$ solute
cluster, as defined in Eq.~(\ref{cluster-var}) but with the actual
equilibrium ratios K$_n$ utilized.  This last form is more elegant
than the previous expressions for the interaction contribution to the
solute chemical potential.  For simulation purposes, however, it is
unlikely to be of direct practical use because $x_0$ will be difficult
to determine with statistical precision from simulation observations.
A virtue of the earlier result in Eq.~(\ref{qca.02}) was its
constructive character, which emphasized how these hydration problems
could be broken down into smaller component problems that individually
might be more susceptible to approximation.  Nevertheless, this
quasi-chemical rule is a new, general, and strikingly simple
expression.

Of course, the $x_0$ indicated in this quasi-chemical rule,
Eq.~(\ref{gqca}), must still be well-defined, therefore the {\em
recognition} problem that was central to the technical discussion
above is still essential.  Nevertheless, the flexibility offered by
the definition of the bonding indicator functions, $b_{\sigma j}$,
should be used in designing quasi-chemical approximations.  The
quasi-chemical rule expresses the compromise that is sought in that
design.  If those bonding regions are defined to be large, the average
$\langle {e^{-\beta \Delta U}}\prod{}_j {(1+f_{\sigma j})} \rangle_0$
should be simple because macroscopic approximations should suffice.
But then $x_0$ would be correspondingly difficult to determine.  In
contrast, if those bonding regions are aggressively defined to be
small, the determination of $x_0$ would be simpler, but then the
exclusion average becomes nearly as difficult as the original problem.
A compromise should be sought that makes each of these component
problems manageable.

\subsection*{Direct Derivation of the Quasi-chemical Rule}
The quasi-chemical rule, Eq.~(\ref{gqca}), is so simple that a direct
derivation is called for.  In view of the potential distribution
theorem in Eq.~(\ref{widom}), a more succinct statement of the
quasi-chemical rule would be
\begin{eqnarray}
x_0 = { \langle {e^{-\beta \Delta U}}\prod\limits_j {(1+f_{\sigma
j})} \rangle_0 \over \left\langle {e^{-\beta \Delta U}}
\right\rangle_0} .
\label{ratio}
\end{eqnarray}
The two indicated averages involve the same sample and the same
normalizing denominators.  The ratio is the average
\begin{eqnarray}
x_0 = \left\langle \prod\limits_j {(1+f_{\sigma j})} \right\rangle
\label{x0}
\end{eqnarray}

\noindent
which includes the solute actually present, not as a test particle,
and involves the full solute-solvent interactions.  The quantity
averaged, $\prod{}_j (1+f_{\sigma j})$, is one (1) if no solvent
molecule penetrates the defined inner shell and zero (0) otherwise.
Thus, $x_0$ of Eq.~(\ref{x0}) is indeed the fraction of Li$^+$ ions
with zero inner shell neighbors.  This identification together with
the potential distribution theorem of Eq.~(\ref{widom}) then gives a
direct derivation of the quasi-chemical rule, Eq.~(\ref{gqca}).  This
derivation justifies the name quasi-chemical rule because
Eq.~(\ref{gqca}) is a simple, directly proved result that permits
regeneration of all the quasi-chemical approximations discussed
earlier.

Though our earlier discussion noted the analogy of the first term on
the right of Eq.~(\ref{gqca}) with the -$\ln p_0$ of
Eq.~(\ref{eqostate}), it is worthwhile to consider an alternative
analogy.  The $x_0$ of the quasi-chemical rule can be interpreted
probabilistically. This suggests that the information theory tools
that have recently been applied to the analysis of p$_0$
\cite{pratt:99,garde:96,Hummer:96,Hummer:98a,Hummer:98b,Gomez:99,%
Garde:99} might be usefully applied to understanding $x_0$ also.
Nevertheless, those are physically distinct problems.

\section*{An Explicit-Implicit Solvation Model}
Next we drive specifically toward a simulation procedure that permits
treatment of the bulk of the solution in an {\em implicit solvent}
fashion.  As a beginning we note some broad points relevant to this
goal.  We do not intend to eliminate all the solvent molecules.  It is
clear that in many cases the solvent water molecules participate
directly and individually on a molecular basis. In the interest of
physical directness of the description, those molecules shouldn't be
eliminated.  The models we seek are then {\em explicit-implicit}
hydration models.  With a few important water molecules explicitly in
the calculation, we will be satisfied with rough theories, such as
dielectric models of the aqueous solution, for the solution more
distant from the explicit action.

Another important point is that we want to carefully limit the
possibilities for explicit water molecules.  This is more than merely
an economic consideration.  We want the thermal distributions of those
explicit water molecules to be simple because it is the entropic
aspects of these problems that are the least clear.  Untempered
structural fluctuations would be an undesirable feature of the
implicit solvent models that we seek.  The introduction of
complications intended to make implicit solvent models more realistic,
such as flexible external boundaries, can make more difficult the
theoretical issues and the problem generally.  The models we envision
will be successful to the extent that only a few explicit water
molecules with simple possibilities for distribution must be treated
and when rough theories will be satisfactory for the implicit solvent
portions of the system.  Our final point here is that we want to embed
these models in a theoretical structure that can be used to analyze
and test the limitations of the inevitable approximations.

\begin{figure}[b]
\centerline{\epsfig{file=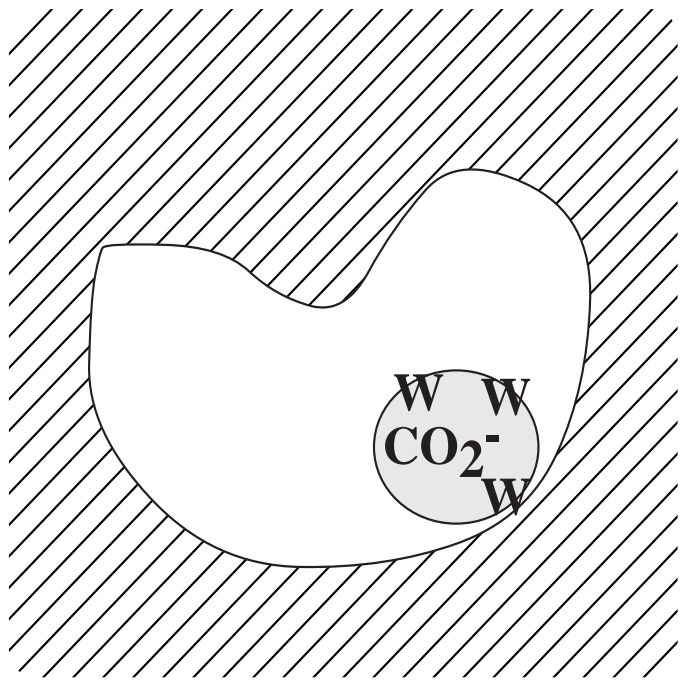,height=3.5in,width=3.5in}}
\vspace{10pt}
\caption{How the quasi-chemical approach might provide an 
explicit-implicit model of macromolecule hydration.  The proximal
volume is defined about a hydrophilic (carboxylate) side chain of a
protein.  The `W's indicate water molecules that might occupy the
defined bonding region.  The material external to the protein plus
ligand complex might be treated by a crude physical model such as a
dielectric continuum model appropriate for longer-ranged and weaker
interactions.}
\label{fig1}
\end{figure}

The idea for how to use the quasi-chemical development of
Eq.~(\ref{gcresults}) as an {\em explicit-implicit} solvation model is
suggested by Fig.~\ref{fig1}.  Proximal volumes are defined around
sites of the macromolecule that require specific care in treatment of
neighboring water molecules.  Referring to the formal treatment above,
this definition implies specification of the bonding functions
$b_{i\sigma}$, and that definition does not require that the protein
structure be rigidly constrained.  As the theory is applied,
therefore, the protein structure can respond to specific features of
the complexation.  Since the volumes permitted for specific hydration
are limited and sharply defined, the possibilities for problematic
fluctuations of the structure of the complex are limited.
Furthermore, the possibilities for a troublesome broad distribution of
ligand occupancies are also limited.

In these complicated settings, it is worthwhile to notice that the
theoretical development can be expressed in a way that is consistent
with understood simulation techniques.  We consider a specific
$n^{th}$ term of the sum of Eq.~(\ref{gcresults}) and then the
inner-most average expressed there:

\begin{eqnarray}
s_{\sigma W_n}{}^{(0)}({\bf 1+n})p_0\left\langle {e^{-\beta \Delta U}}
\right\rangle _0^*z^nq_{\sigma W_n}=e^{-\beta {\mathcal F}_{\sigma W_n}({\bf 1+n})}\left( {{{z^n}
\over {n!}}} \right) .
\end{eqnarray}

\noindent This equation introduces the conformational free energy
${\mathcal F}_{\sigma W_{n}}({\bf 1+n})$ associated with a geometry of the
solute and $n$ specific solvent (water) molecules within the defined
proximal volume.  The factor $n!$ on the right arises because
$n!\,q_{\sigma W_n}$ is the normalizing feature of $s_{\sigma
W_n}{}^{(0)}({\bf 1+n})$ involved in Eq.~(\ref{nth}).  Our
quasi-chemical master formula, Eq.~(\ref{gcresults}), can then be
expressed as\footnote{ $\rho/z$ is a standard combination of variables
for reasons that Eq.~(\protect\ref{widom}) makes clear.  It is also
common to employ the notation $\zeta_\alpha = z_\alpha q_\alpha /V$ so
that $\rho_\alpha/\zeta_\alpha \rightarrow$ 1 when interactions are
negligible.}

\begin{eqnarray}
{\rho_\sigma \over z_\sigma} = V^{-1}\sum\limits_{n \ge 0} {\left(
{{{z^n} \over {n!}}} \right)Tr_n\left( {e^{-\beta {\mathcal F}_{\sigma
W_n}}}
\right)} .
\label{qcgc}
\end{eqnarray}

\noindent
$Tr_n \left( \ldots \right)$ indicates the trace as it might appear,
for example, in the conventional evaluation of a canonical partition
function.  That term evaluates to $n!\,q_{\sigma W_n}$ --- the
`configuration integral' \cite{hill:config_int} for the isolated
cluster --- in the case that the external-cluster interactions are
negligible. The formula Eq.~(\ref{qcgc}) suggests a grand canonical
ensemble calculation but with some primitive and essential
differences.  Firstly, the volume of the system is attached to the
solute and, in fact, this volume can change as the solute changes
conformation.  Secondly, the function appearing where the potential
energy would appear is the conformational free energy
${\mathcal F}_{\sigma W_n}({\bf 1+n})$.  This depends on the
thermodynamic state of the solvent, that is, on $z$ and $\beta$.  When
$z$=0, only the n=0 term of Eq.~(\ref{qcgc}) is non-zero and then
Eq.~(\ref{qcgc}) is trivially verified with $\beta\Delta \mu _\sigma$=0.
Finally, the thermodynamic identification of this partition function is
the combination on the left side of Eq.~(\ref{qcgc}).

Despite these important differences, the structural similarity of
Eq.~(\ref{qcgc}) with a grand canonical partition function is
remarkable.  Calculations that used this approach would be built upon
a model for ${\mathcal F}_{\sigma W_n}({\bf 1+n})$, which included
physical approximations for the extra-cluster contributions, and on
the simulation algorithms for calculations on grand canonical
ensembles.  Grand canonical simulation calculations are less routine
than other types of simulations and probably more demanding when
applied to condensed phases.  If grand canonical simulations were
prohibative in a particular case because of a lack of facility of
molecule exchange, then these approaches might also be prohibitive
because of that.  Practical points of difference might help, however.
The present development is designed so that the number of ligand
molecules needn't be great.  Additionally, the conformational changes
of the solute can change the defined proximal volume, changing the
local ligand binding characteristics either to squeeze out ligands or
to open space for insertion of new ligands.

Again we emphasize the idea that if the proximal volumes are defined
to be aggressively small then the evaluation of the partition
functions indicated in Eq.~(\ref{qcgc}) should be simpler. In the most
favorable case, the indicated partition functions might be evaluated
by a maximum term and gaussian distribution procedure.  In that case,
the computational work to exploit this proposal is only slightly
greater than a determination of an optimum hydration structure for the
ligands. Roughly put, the theory provides a justification for treating
a small set of waters as part of the solute molecule under study.  The
theory then suggests consideration of the `chemical' reaction for
isolated reaction species.  According to this approach, the study of
the ligands and of the complex, then inclusion of a simple implicit
solvent model permits an inference of the basic hydration free energy
for the solute alone.

\section*{Discussion}
In the area of computational theory for aqueous solution chemistry and
biochemistry, there seems to be a well developed folklore that
judicious inclusion of a pivotal few water molecules is typically the
important next step beyond dielectric continuum implicit solvent
models.  The quasi-chemical approach developed here is the statistical
mechanical theory for how to do that properly for solution
thermodynamics.

\subsection*{Implicit Solvent Models}
One goal of an implicit solvent model is to increase computational
scope.  Otherwise inaccessible problems might become accessible to
study.  Or, perhaps, hundreds of instances might be checked with
implicit solvent models where only one instance could be examined if
sufficient water must be explicitly included.  This increased scope is
accomplished, in concept at least, by reducing the amount of solvent
that must be explicitly tracked in a simulation calculation.  The
prices to be paid for this reduction are the approximations that must
be accepted and the increased complexity of the implicit calculation.
How the advantages and complications balance in particular cases
typically has been unclear.

Nevertheless, there are other potential advantages of implicit solvent
models and the chief of these is the same advantage offered by
approximate physical theories.  Scientific use of such theories serves
to establish simplified concepts that are valid and valuable where the
goal is understanding as well as predicting.  The physical idea
implicit in the developments above is that a handful of proximal water
molecules are decisive for chemistry and biochemistry in aqueous
solution.  Even for such extended molecular events as protein folding,
it is a valuable hypothesis that a small number of water molecules,
perhaps hundreds rather than a hundred times that, play a decisive
role.

The quasi-chemical factors of $\rho_{H_2O}{}^n$ of
Eqs.~(\ref{eqostate}) and (\ref{qca.01}) are the most basic feature of
the environment mean-field operating on the solvent molecules included
explicitly.  This identification is a basic difference from other
implicit solvent models and suggests the possibility of aggressively
pushing the numbers of explicitly included solvent molecules down to
minimal chemically reasonable values.

\subsection*{Dielectric Models and Electronic Structure Calculations 
for Solution Thermodynamics} Explicit treatment of water molecules
that are near-neighbors of the solutes should
mitigate\cite{Hummer:97c} the awkward, not always well-defined
invocations of ``dielectric saturation and electrostriction.''  In
this way, quasi-chemical developments also address lingering
foundational problems with dielectric models in applications to
solution chemistry.  Parameterization of dielectric models to describe
variations of hydration free energies with temperature, pressure,
composition, and solute
conformation\cite{Pratt:94a,Tawa:94,Tawa:95,Pratt:97,Hummer:98} are as
relevant as the foundational issues.  For the quasi-chemical models
used here, some empirical parameterization is still required for the
ligands that form the exterior of the complex.  In the Li$^+$ example,
this means radii parameters are required for the water molecules
involved.  However, the results are insensitive to radii
parameters required for the solute Li$^+$ ion; and furthermore, as the
theoretical development emphasizes, the product of such efforts is the
thermodynamic characteristics of the {\em solute}, not the ligands.
The development above has emphasized that much of the thermodynamic
state dependence of hydration free energies is explicitly available in
quasi-chemical approaches.

Similarly, phenomena involving the electronic structure of the solute
are naturally, though approximately, included in quasi-chemical
theories.  Examples of such phenomena are nuclear versus electronic
time scale issues, dispersion interactions and electron correlation
more generally, the orthogonality or `charge-leakage' issues that are
currently discussed for dielectric models, and solute-solvent charge
transfer that was observed for the Li$^+$ example, see Table~\ref{T_1}.  They may be
excluded in unvarnished dielectric models, or included in only {\em ad
hoc} ways.  Since it is the molecular and thermodynamic properties of
the {\em solute} that is the goal of the quasi-chemical theories, the
phenomena listed above are reasonably included for the solute
properties.

\section*{Conclusions}
This paper has given an extended development of the quasi-chemical
approach to computational solution chemistry and biochemistry and
discussed the relevance of this development to problems of implicit
solvent models of the structural molecular biology.  The new implicit
solvent model presented here is, more accurately, an explicit-implicit
model applicable to the statistical thermodynamics of solutions. The
physical idea for the model is that a small subset of solvent
molecules are decisive for the statistical thermodynamics and require
explicit treatment, whereas the remaining solvent molecules are less
important and may be treated implicitly.

The formal derivation of the quasi-chemical formula depends on the
well-known potential distribution theorem aided by a clustering
analysis that serves as a counting device.  In the process of the
derivation, it is shown how the view of the system is transformed from
a single solute in solution to a solute complexed to solvent ligand
molecules interacting with the extra-complex solution. The
thermodynamic derivation and the quasi-chemical rule that results,
Eq.~(\ref{gqca}), are new.

To illustrate the quasi-chemical approach, the hydration thermodynamic
properties of Li$^+$(aq) are calculated. The problem is framed as a
series of ion hydration reactions involving the isolated ion,
water molecule ligands, and ion-ligand complexes.  Application of an
implicit solvent model to the species, the standard dielectric continuum
model in this case, accounts for interactions with the extra-complex
solvent molecules. The results predict that Li$^+$ in liquid water is
surrounded by four (4) inner shell water molecule ligands with a
chemical potential comparable to experimental results. Since the
explicit pressure and temperature dependencies of the chemical potential
are analytically available in the quasi-chemical formula, results for
the partial molar volume and entropy of Li$^+$(aq) are explicitly
available. The evaluations of the partial molar volume and partial molar
entropy of the Li$^+$(aq) give a snapshot of the status of molecular
theory for these properties.

\section*{Acknowledgment}  This research is supported by the
Department of Energy, under contract W-7405-ENG-36 and the LDRD
program at Los Alamos.  The model here proposed was motivated by May
1998 CECAM Workshop ``Implicit solvent models for biomolecular
simulations,'' organized by Roux and Simonson\cite{Roux:99a,Roux:99b}.
The further theoretical development in December 1998 was facilitated
by M. E. Paulaitis and the hospitality of the Johns Hopkins University
Chemistry Engineering department.

\end{document}